\documentclass[review]{article}
\usepackage{setspace}
\usepackage{epsfig}
\usepackage[multiple]{footmisc}
\usepackage{amsmath}
\usepackage{verbatim}
\usepackage{graphicx}
\usepackage{amssymb}
\usepackage{color}
\usepackage{subfig}
\usepackage{lineno}
\usepackage[top=2cm, bottom=2cm, left=2.5cm, right=2.5cm]{geometry}

\begin{document}
\title{Timing and Charge measurement of single gap Resistive Plate Chamber Detectors for $INO-ICAL$ Experiment}

\author{ Ankit Gaur\footnote{Corresponding author: ankitphysics09@gmail.com} , Ashok Kumar, Md. Naimuddin
\\
Department of Physics and Astrophysics, University of Delhi,\\ Delhi 110007, India. }

\maketitle 

\begin{abstract}
{The recently approved India-based Neutrino Observatory will use the world's largest magnet to study atmospheric muon neutrinos. The 50 kiloton Iron Calorimeter consists of iron alternating with single-gap resistive plate chambers. A uniform magnetic field of $\sim$1.5 T is produced in the iron using toroidal-shaped copper coils. Muon neutrinos interact with the iron target to produce charged muons, which are detected by the resistive plate chambers, and tracked using orthogonal pick up strips. Timing information for each layer is used to discriminate between upward and downward traveling muons. The design of the readout electronics for the detector depends critically on an accurate model of the charge induced by the muons, and the dependence on bias voltages. In this paper, we present timing and charge response measurements using prototype detectors under different operating conditions. We also report the effect of varying gas mixture, particularly $SF_6$, on the timing response.}

\end{abstract}
\vspace{1cm}
~~~~~~~~ {Keywords: ICAL, Timing, Gaseous Detector, RPC }

 \newpage
%keywords{ICAL, Timing, Gaseous Detector, RPC}

%\begin{document}

\section {Introduction}
\label{sec:intro}

In the Standard Model of particle physics, neutrinos are leptons with zero-charge, so they do not participate in either electromagnetic or strong interactions.  As leptons, they have spin one-half, and come in three flavors (electron, muon, or tau) named after the charged-lepton flavor with which they partake in weak interactions.  Once assumed to have zero-mass, the observation of neutrino flavor-oscillation revealed that at least some flavors of neutrino must have a tiny, but non-zero mass.   A concerted worldwide effort to study neutrino oscillations and their masses \cite{kam}-\cite{cap} has made tremendous progress unraveling the properties of these most elusive of particles.  However, there remain many unresolved questions about the nature of neutrinos.  The Indian physics community is considering the construction of a world-class underground neutrino facility, the India-based Neutrino Observatory (INO) \cite{ino}, in order to answer some of the most important questions about the nature of neutrinos including their mass hierarchy, the extent of CP violation in their interactions, and whether or not they are Majorana particles (i.e. their own anti-particles).

The INO facility will host multiple experiments including the Iron Calorimeter (ICAL) detector, which is optimized for the study of atmospheric neutrinos. 
The ICAL detector, will consist of three $17~{\rm kt}$ modules each measuring $16~{\rm m}\times16~{\rm m}\times14.5~{\rm m}$ and containing 151 iron plates, $56~\rm{mm}$ each, interleaved with Resistive Plate Chambers (RPCs).

The iron plates provide sufficient mass to cause passing atmospheric neutrinos to interact,  while the RPC detectors provide tracking and  fast timing measurements.  The excellent timing and spatial resolution \cite{san}-\cite{cms} of RPC detectors allows for highly accurate time of flight measurement \cite{blanco}-\cite{blanco2}. The ICAL detector will distinguish between upward and downward traveling muon neutrinos in order to enhance the sensitivity of its physics program. As neutrinos travel at close to the speed of light, a timing resolution of the order of a few tens of nanoseconds is needed to infer the direction of travel from the arrival time at opposite ends of the ICAL detector. In addition to the timing performance, an accurate model for the charge content of pulses in the ICAL detector is needed to design the readout electronics.   In this paper, we report on both the timing resolution and charge measurements on prototypes of the ICAL RPC detectors under a variety of operating conditions.

\section {The ICAL RPC Detector}
\label{sec:ical}

RPC detectors are characterized by a high efficiency and a fast response time, achieved at relatively low construction cost.  Each RPC consists of two resistive plates, an anode and a cathode, 
typically fabricated from either bakelite or glass, separated by a gap filled with a gaseous mixture.  During operation, a high voltage applied across the electrodes induces an electric field within the gaseous volume.  Charged particles traversing the gap ionize the gas, producing electron-ion pairs along their trajectory.  The electron-ion pairs accelerate, in opposite directions, under the influence of the electric field, reaching sufficient kinetic energy to further ionize the gas.  At a sufficiently high voltage this process produces an avalanche which significantly amplifies the charge that eventually reaches the resistive plates.  The readout electronics converts the charge that reaches each plate into electrical signals which encode the time and charge distribution of each pulse.  More details on the operation of RPC detectors can be found in \cite{san}, while the detailed geometry and construction procedures are in ~\cite{daljit}-\cite{bakelite}. 

The massive $50~{\rm kt}$ ICAL sampling Calorimeter requires 28,000 RPC detectors, the surface of which is $2~\rm{m} \times 2~\rm{m}$. These RPCs are constructed with 3 mm float glass and read out by 2.8 cm wide strip, made up of copper. For efficient and reliable operation of the RPCs under a variety of environmental conditions, a dedicated $R\&D$ effort is crucial for optimizing the operating parameters. In this study, we examine the effect of operational parameters on the timing resolution and charge spectra.  The performance study of RPCs have also been carried out under different gaseous mixtures, details of which are provided in section~\ref{sec:performance}.

\section{Experimental setup}
\label{sec:setup}

\begin{figure}[htbp]
\centering
\includegraphics[height=7cm,width=10cm]{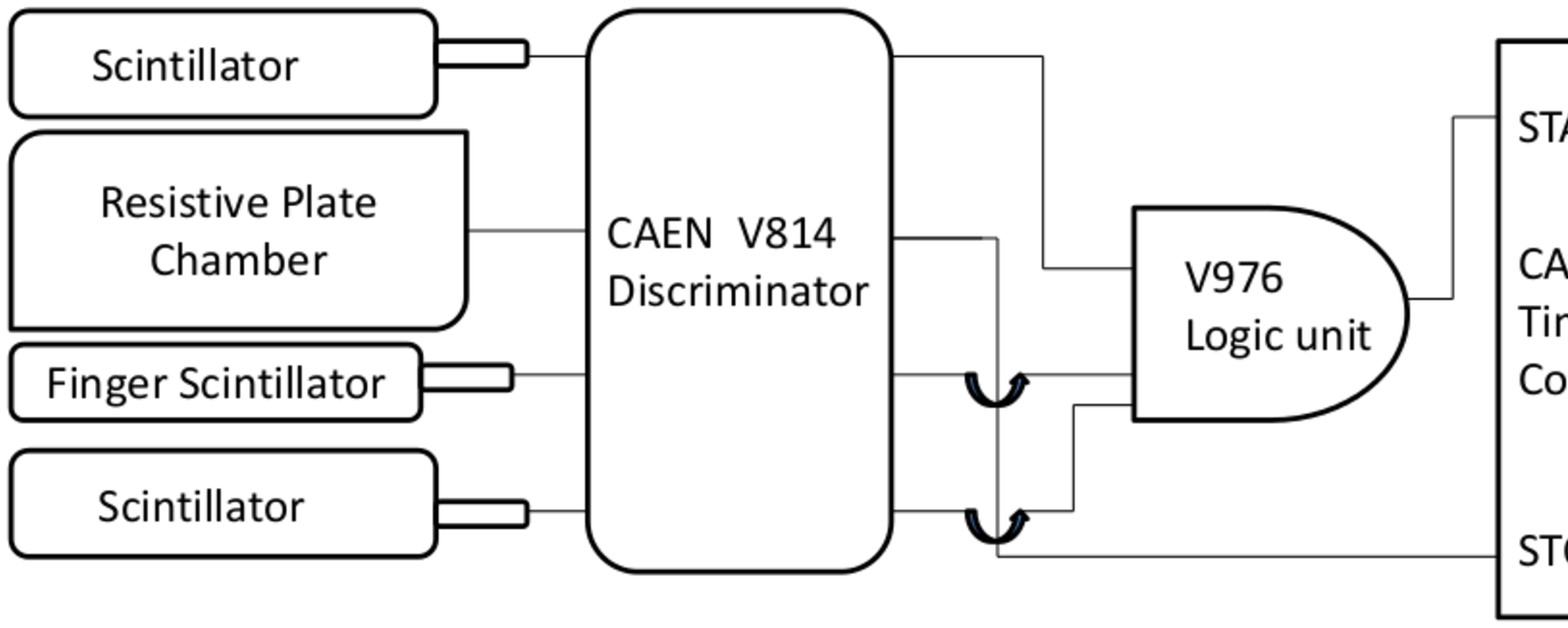}
\caption{Schematic diagram of the experimental set-up used for timing response measurements.}
\label{fig:tdc}
\end{figure}

\begin{figure}[htbp]
\begin{minipage}{\linewidth}
      \centering
      \begin{minipage}{0.49\linewidth}
          \includegraphics[width=9cm,height=6cm]{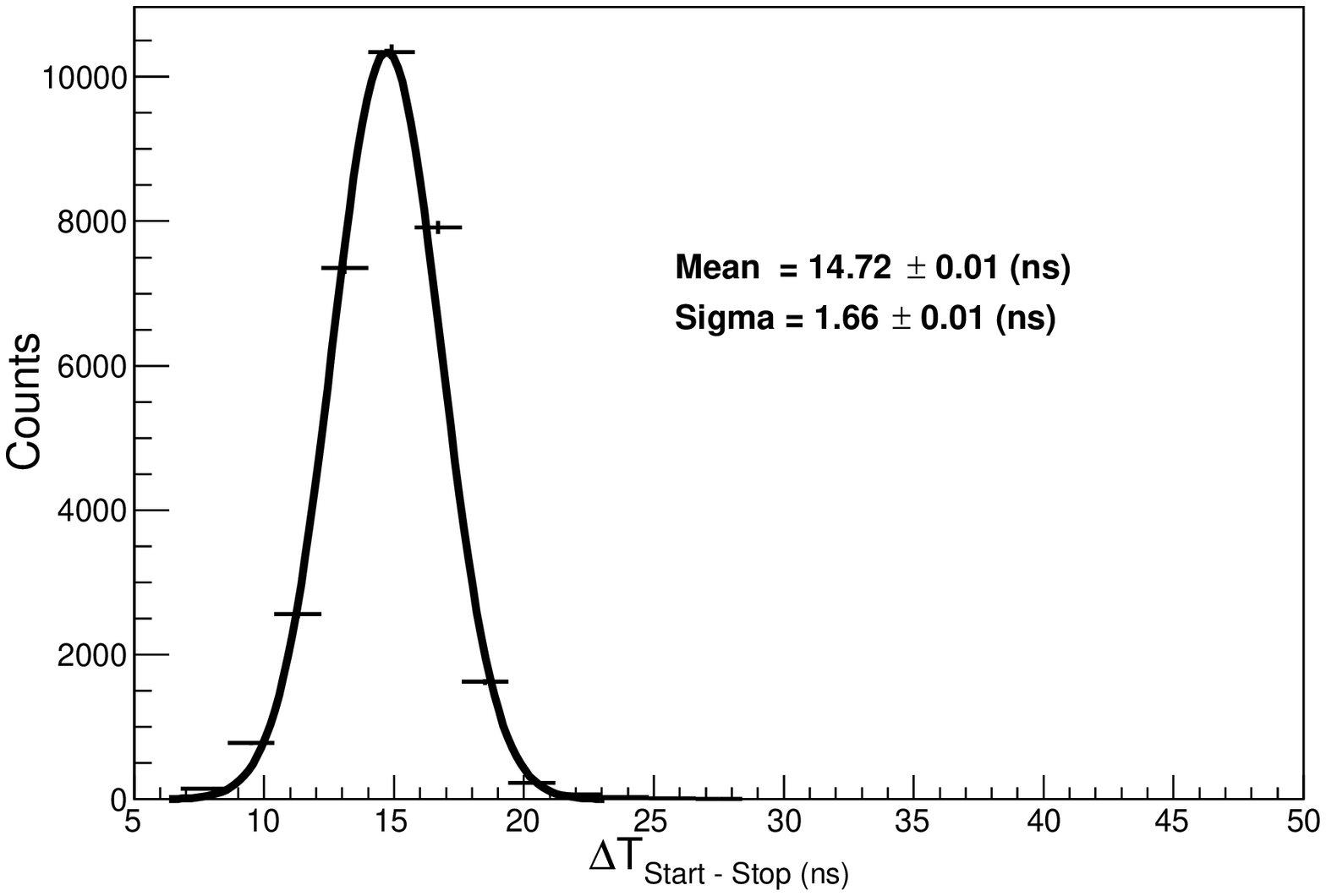}
               \end{minipage}
      \begin{minipage}{0.49\linewidth}
%\centering
            \includegraphics[width=9cm,height=6cm]{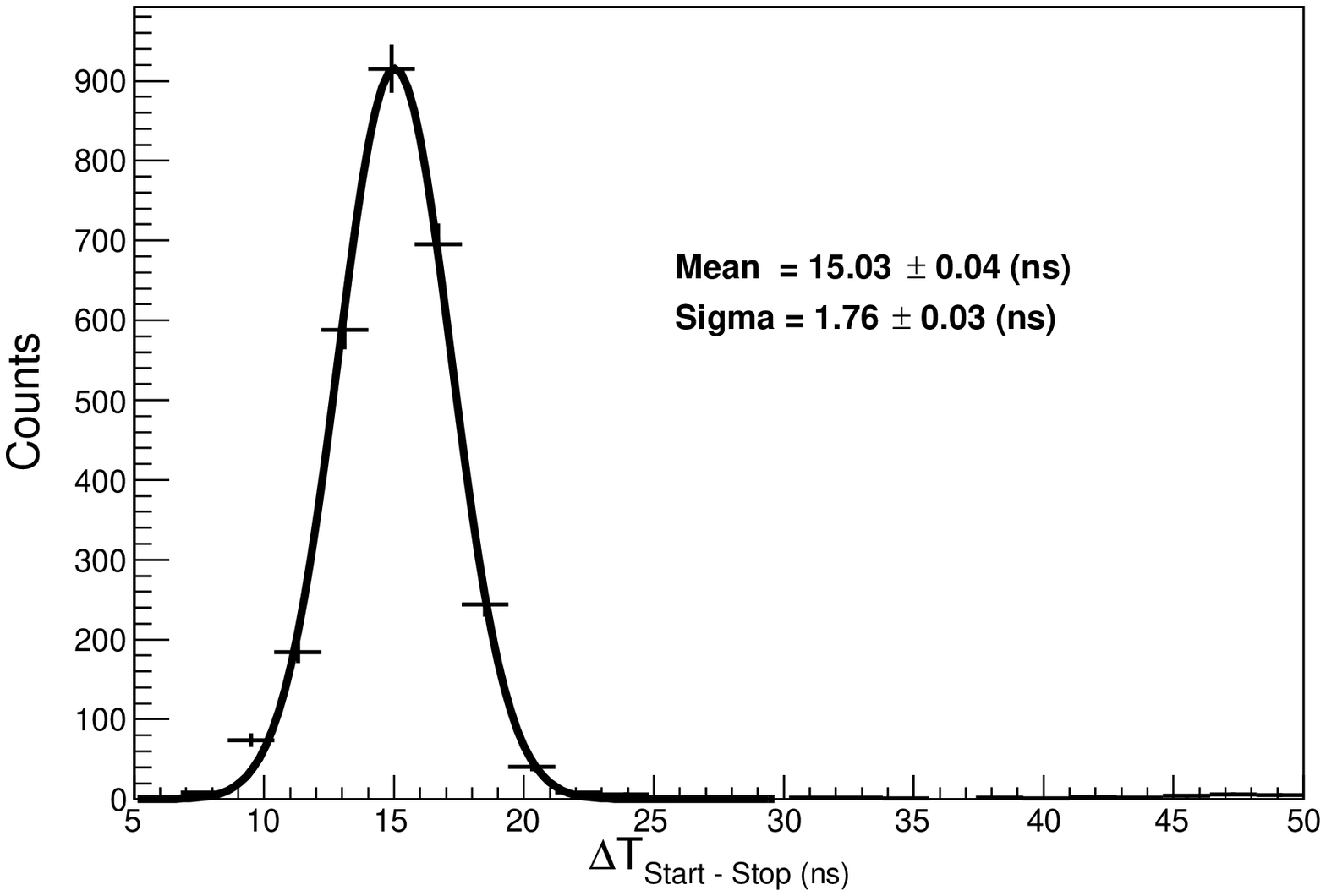}
             \end{minipage}
                    \end{minipage}
\caption{Time distribution for Saint Gobain (left) and Asahi (right) RPC detectors at a bias voltage of 10.6 kV, $SF_6$ concentration of 0.3\%, and at discriminator threshold of 50 mV. }
          \label{fig:timing}
 \end{figure}

\begin{figure}[htbp]
\centering
\includegraphics[height=7cm,width=10cm]{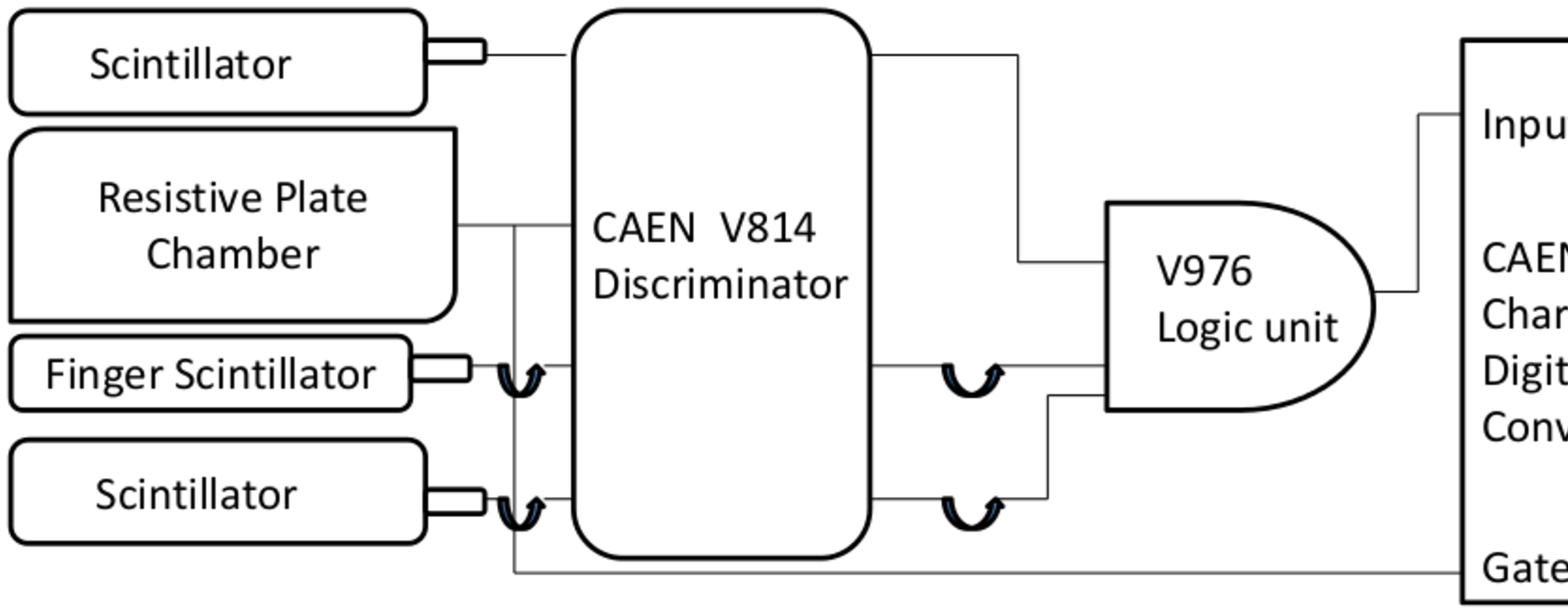}
\caption{ Schematic diagram of the experimental set-up used for charge response measurements from RPC detector.}
\label{fig:qdc}
\end{figure}

\begin{figure}[htbp]
\begin{minipage}{\linewidth}
      \centering
      \begin{minipage}{0.49\linewidth}
          \includegraphics[width=9cm,height=6cm]{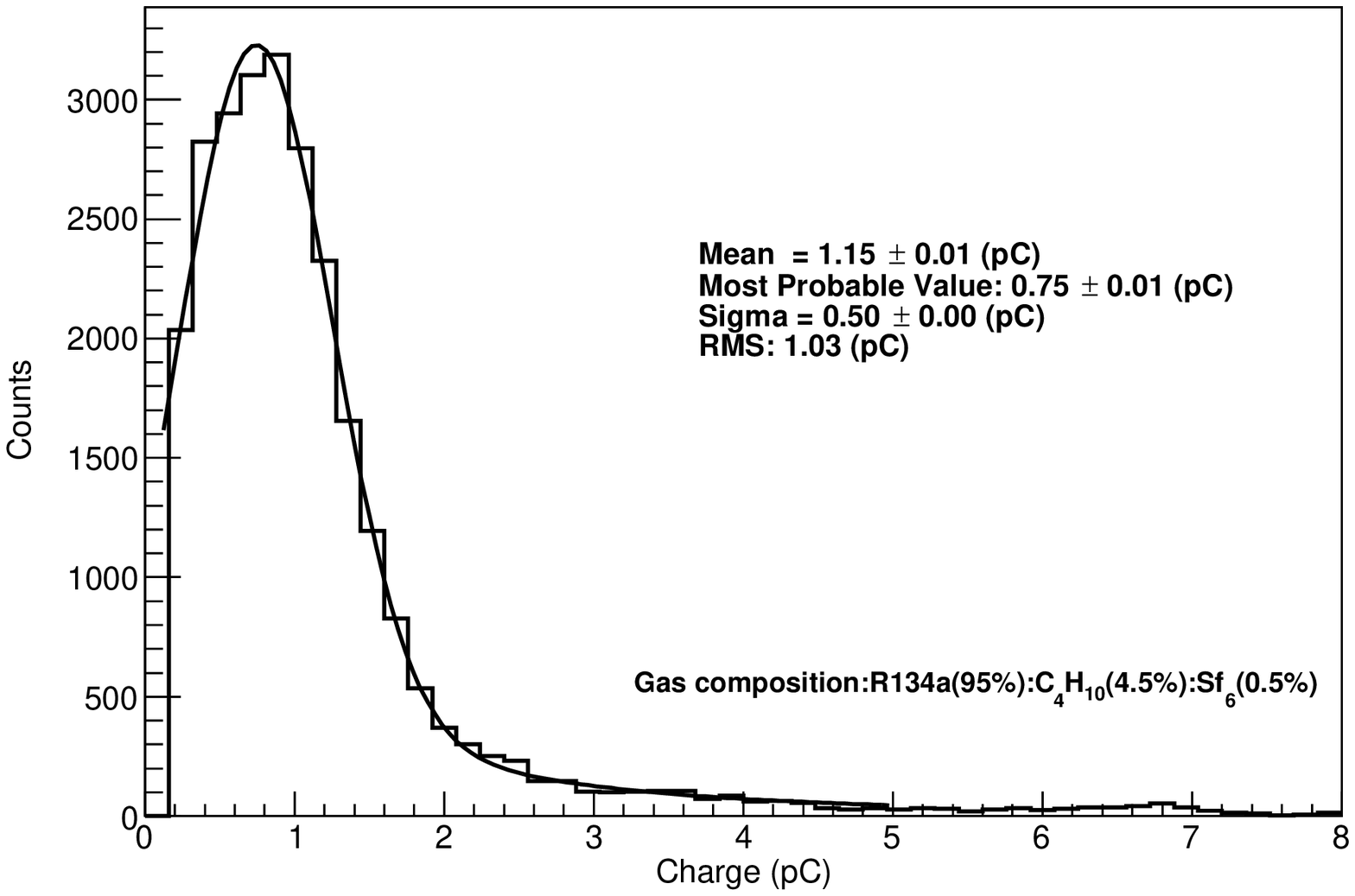}
               \end{minipage}
      \begin{minipage}{0.49\linewidth}
%\centering
            \includegraphics[width=9cm,height=6cm]{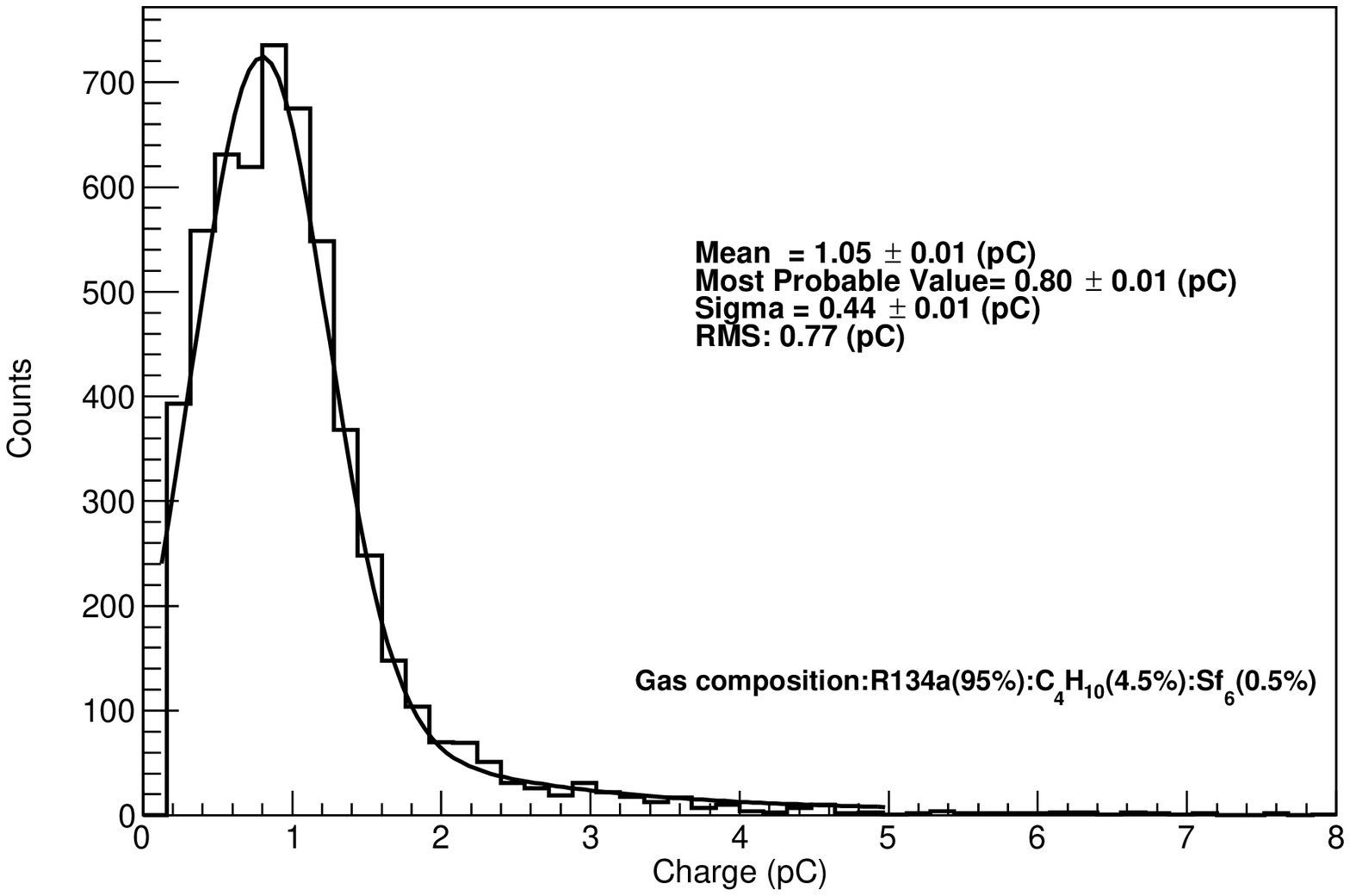}
             \end{minipage}
                    \end{minipage}
\caption{Charge distribution for Saint Gobain (left) and Asahi (right) RPC detectors at a bias voltage of 10.2 kV and $SF_6$ concentration of 0.5\%.}
          \label{fig:charge}
\end{figure}

\begin{figure}[htbp]
\begin{minipage}{\linewidth}
      \centering
      \begin{minipage}{0.49\linewidth}
          \includegraphics[width=9cm,height=6cm]{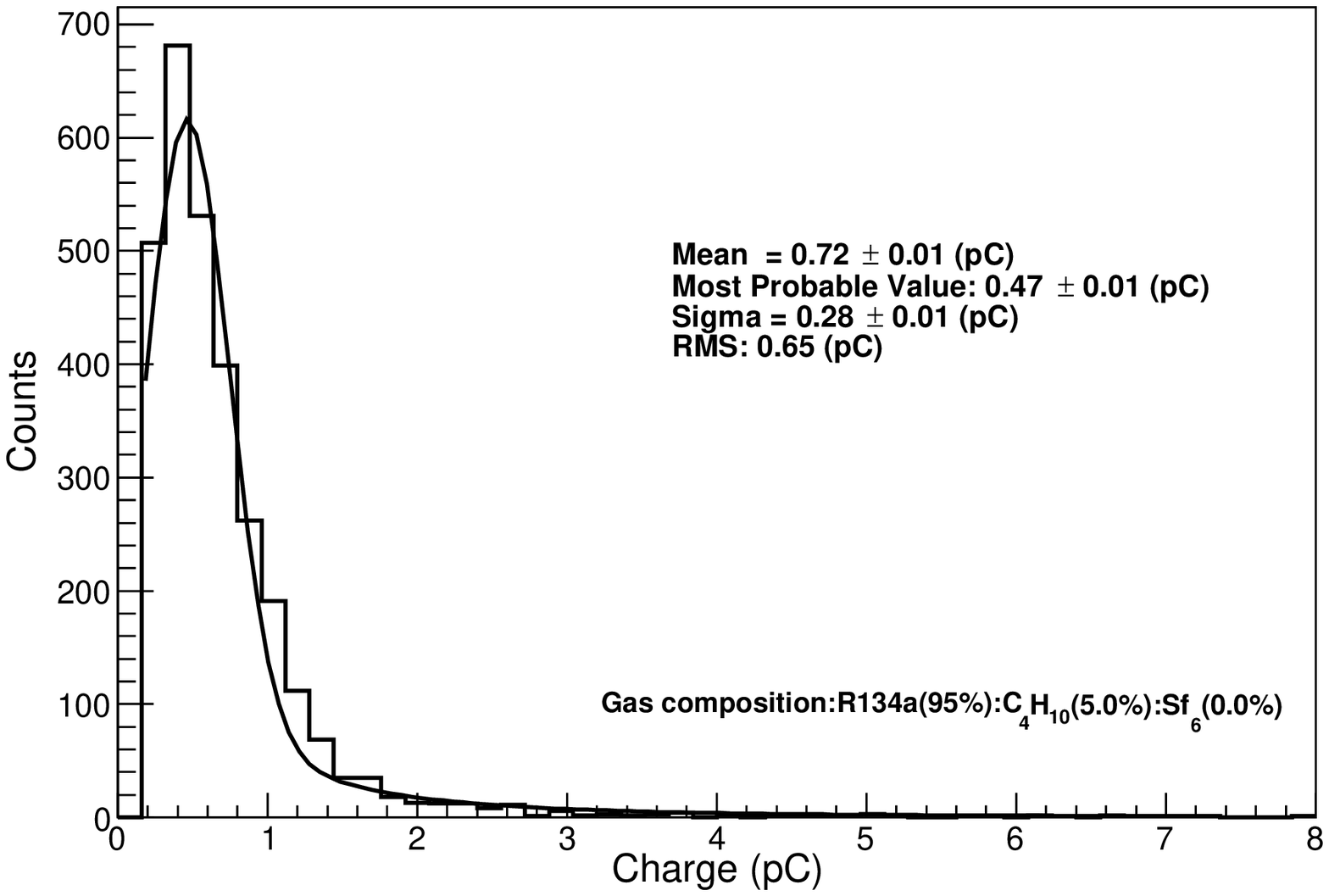}
               \end{minipage}
      \begin{minipage}{0.49\linewidth}
%\centering
            \includegraphics[width=9cm,height=6cm]{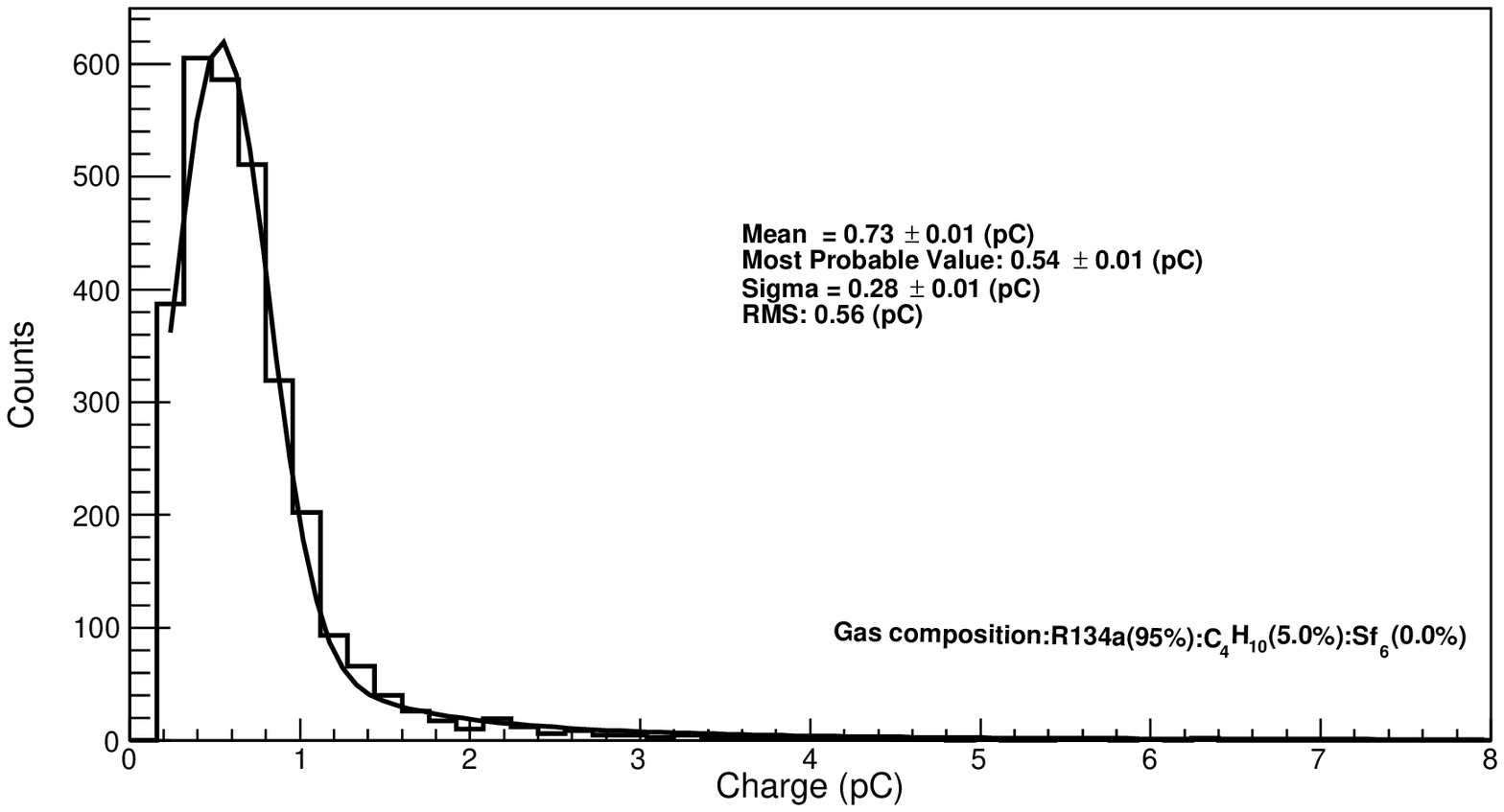}
             \end{minipage}
                    \end{minipage}
\caption{Charge distribution for Saint Gobain (left) and Asahi (right) RPC detectors at a bias voltage of 10.2 kV and $SF_6$ concentration of 0\%.}
          \label{fig:charge1}
\end{figure}

The performance of a prototype ICAL RPC detector was characterized using cosmic ray muons detected with a plastic scintillator hodoscope.   The RPC detector is placed between two large scintillators, each consisting of a polyvinyl toluene (PVT) polymer and instrumented with a Hamammastu H3178-51 Photomultiplier tube (PMT).   An additional thin "finger" scintillator, the size of a single readout strip in the RPC detector allows for more precise location of incoming cosmic ray muons. The analog output pulses from the three scintillators are fed into a CAEN V814 Leading Edge Discriminator with minimum width and pulse heights optimized to reject instrumental noise while maintaining high-efficiency for cosmic ray muons.  In the case of the large scintillators, the pulse width is set at $40~\rm{ns}$ and the pulse height threshold is set at $25~\rm{mV}$, while for finger scintillator the settings are $40~\rm{ns}$ and $50~\rm{mV}$.   
 The discriminator output corresponding to each scintillator is fed into a CAEN V976 logic unit, where a three-way coincidence provides the trigger signal which is used as a START command for a 
 CAEN V775 time-to-digital converter (TDC).  In order to amplify short raw output pulses from the RPC, they are  fed into a preamplifier, followed by an amplifier circuit characterized by a bandwidth of 0.1-1 GHz and a gain of $\sim$60. The output pulses from the amplifier are then fed into a CAEN V814 Leading Edge Discriminator, whose output after providing the appropriate delay with respect to the trigger signal  is utilized as a common STOP command for the  CAEN V775 TDC. The time interval
  between the START and STOP signal is converted into voltage level using the built-in TAC (time to analog converter) feature of the multichannel TDC module.  The output of the TAC sections are multiplexed and subsequently converted by two fast ADC (analog to digital converter) modules. A TDC  produces output which is the absolute time difference between the START and STOP and fluctuation in their value provides the estimate of time resolution. The block diagram for the measurement of timing resolution is shown in  Fig.~\ref{fig:tdc}. We show in  Fig.~\ref{fig:timing}, the distribution of the time difference between the START and the STOP signal. The width of this distribution provides an estimate of timing resolution.
The development of ionization charge and its amplification within the RPC detector depend, among others, on the composition of the gaseous mixture. In the avalanche mode, typical gases Tetrafluoroethane (TFE), isobutane and sulphur hexaflouride are used, yielding raw signals (without preamplifier) of an amplitude of 2-5 mV under appropriate electric fields. The study of charge development due to ionization, resulting in avalanche, and their nature with respect to applied voltage gives a panoramic view of avalanche to streamer transition. A charge to digital converter (QDC) has been used for the study of charge produced under various operating conditions. The schematic of the setup for charge measurement is shown in  Fig.~\ref{fig:qdc}. The output pulses (analog) from each scintillator is fed into a  CAEN V814 Leading Edge Discriminator for analog to digital conversion. The discriminated outputs of the scintillators are sent to CAEN V976 logic unit to obtain GATE pulse for the CAEN V965A QDC, while the analog output of the RPC after providing the appropriate delay with respect to the trigger signal used as an input for the GATE.  Fig.~\ref{fig:charge} and Fig.~\ref{fig:charge1}  show an example of charge distribution for with and without $SF_{6}$  gas mixture and at a particular bias voltage of 10.2 kV. 
                       

\section{Performance study with varying gas mixtures} 
\label{sec:performance}

The RPC detector performance is strongly linked to the gas mixture it uses. It has been shown that the addition of sulphur hexafluoride ($\rm{SF}_6$) to the mixture in the avalanche mode of operation, results in the reduction of charge produced inside the detector following the passage of charge particle along with the suppression of streamers~\cite{abe2}. Although the exact mechanism is not yet well understood, the electron affinity of  $\rm{SF}_6$ could explain the charge reduction by the absorption of part of the produced electrons.
The suppression of streamers results in streamer free operation across a large voltage range ( i.e., in the plateau region) and more stable operation at higher voltages.   As the total charge produced in the gas during each event is also reduced, the presence of $\rm{SF}_6$ also slows the aging of detector electrodes, a crucial consideration for long term stable operation of an RPC.
However, this stable operation does come at the cost of a reduced efficiency at low voltages, as the same charge suppression effect leading to increased stability also suppresses small charge production at lower voltages. 

\begin{figure}[htbp]
\begin{minipage}{\linewidth}
%      \centering
      \begin{minipage}{0.49\linewidth}
          \includegraphics[width=9cm,height=6cm]{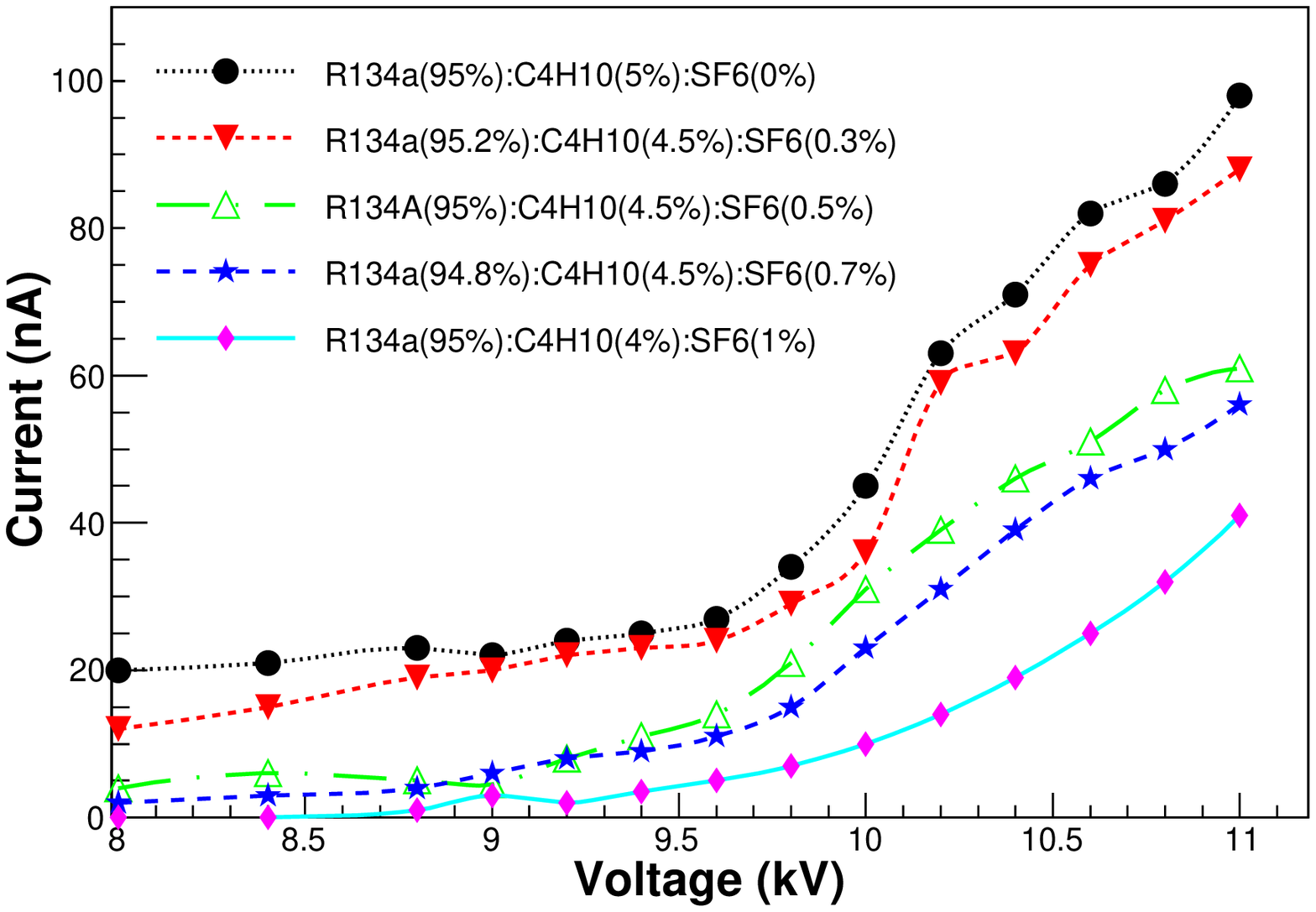}
               \end{minipage}
      \begin{minipage}{0.49\linewidth}
%\centering
            \includegraphics[width=9cm,height=6cm]{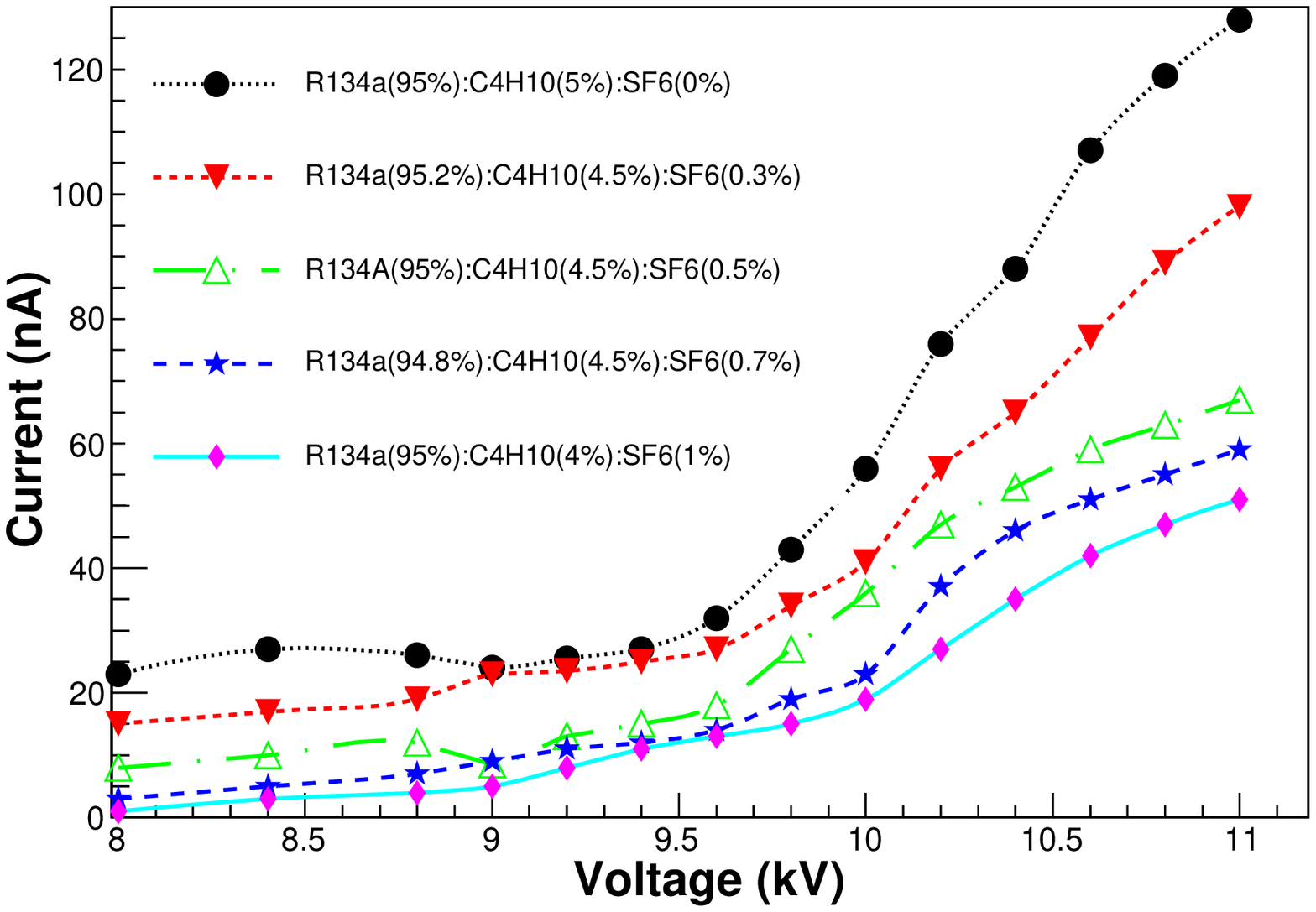}
             \end{minipage}
                    \end{minipage}
\caption{Leakage Current of Saint Gobain (left) and Asahi (right) RPC for all the studied five gas mixtures.}

\label{fig:crnt}
          
 \end{figure}

\begin{figure}[htbp]
\begin{minipage}{\linewidth}
%      \centering
      \begin{minipage}{0.49\linewidth}
          \includegraphics[width=9cm,height=6cm]{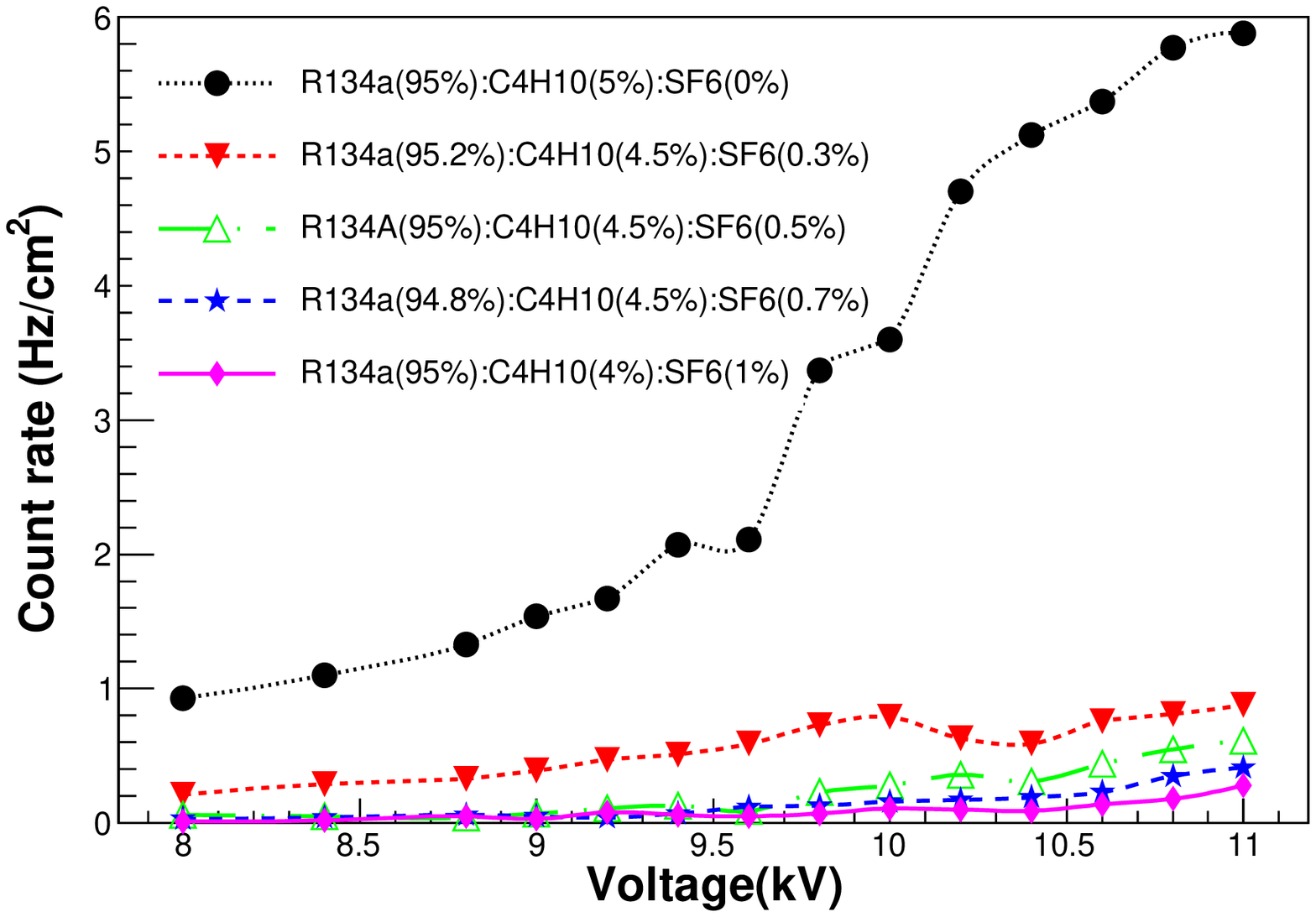}
               \end{minipage}
      \begin{minipage}{0.49\linewidth}
%\centering
            \includegraphics[width=9cm,height=6cm]{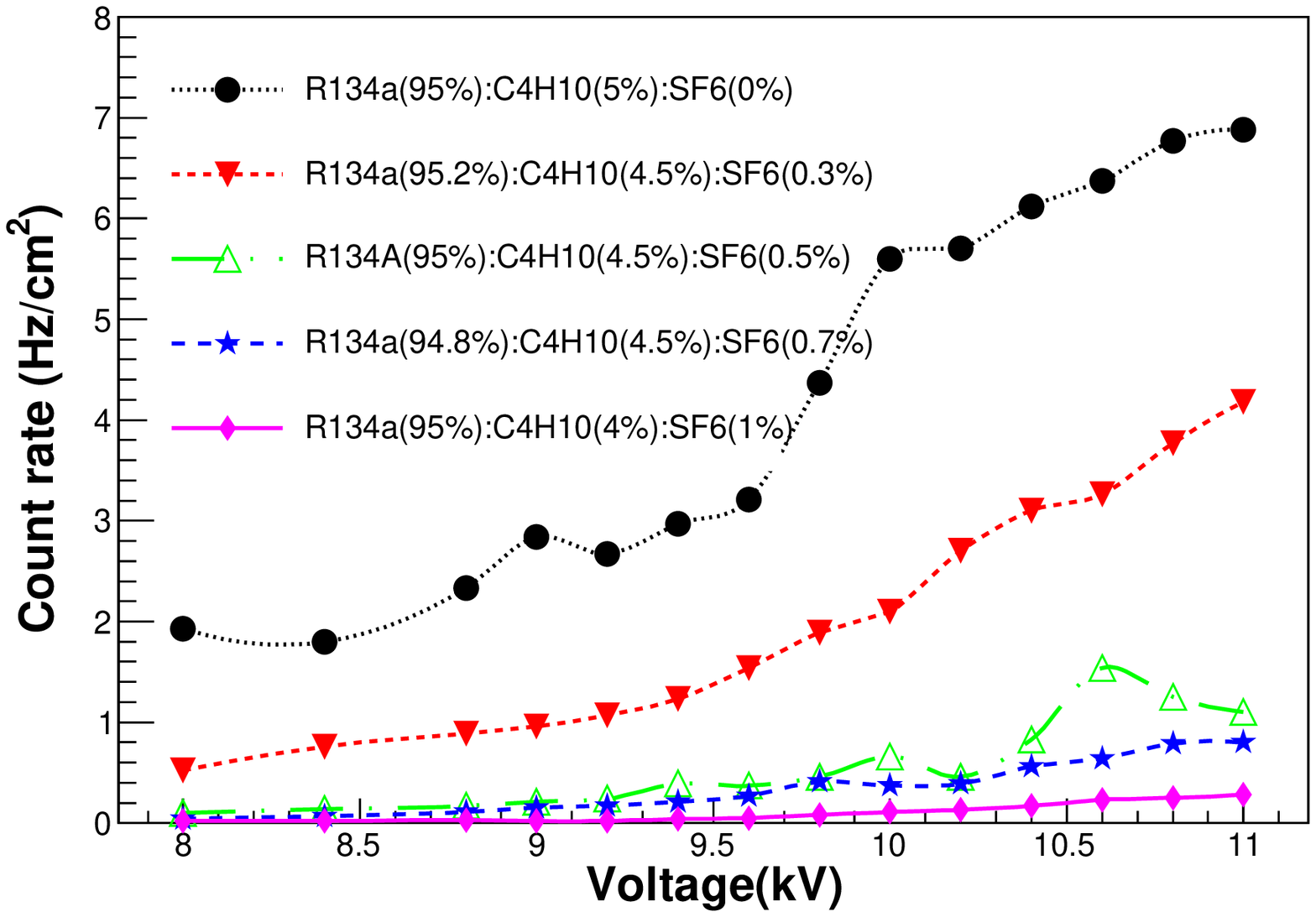}
             \end{minipage}
                    \end{minipage}
\caption{Count rate of Saint Gobain (left) and Asahi (right) RPC for all the studied five gas mixtures.}
\label{fig:noise}
          
 \end{figure}

\begin{figure}[htbp]
\begin{minipage}{\linewidth}
%      \centering
      \begin{minipage}{0.49\linewidth}
          \includegraphics[width=9cm,height=6cm]{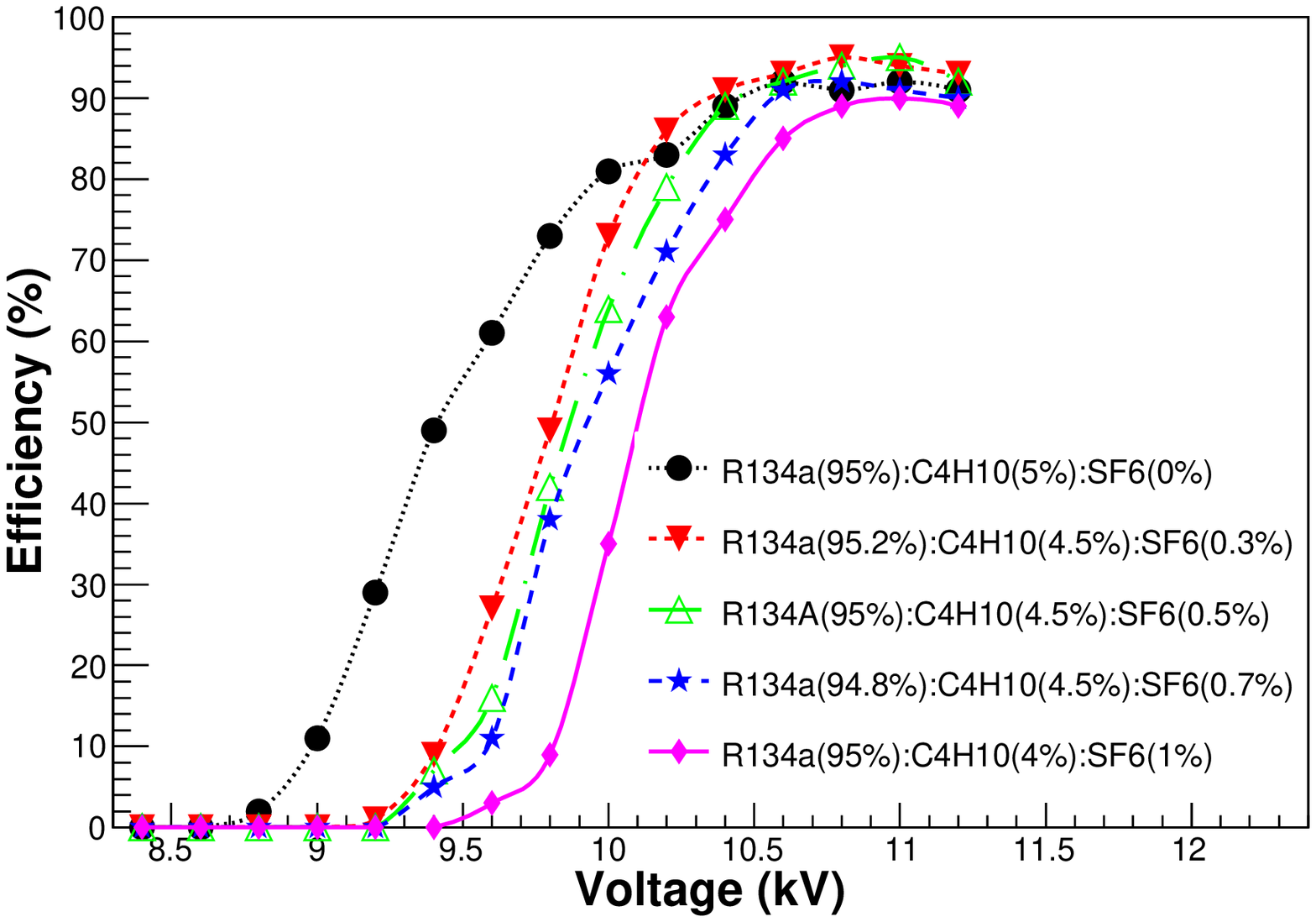}
               \end{minipage}
      \begin{minipage}{0.49\linewidth}
%\centering
            \includegraphics[width=9cm,height=6cm]{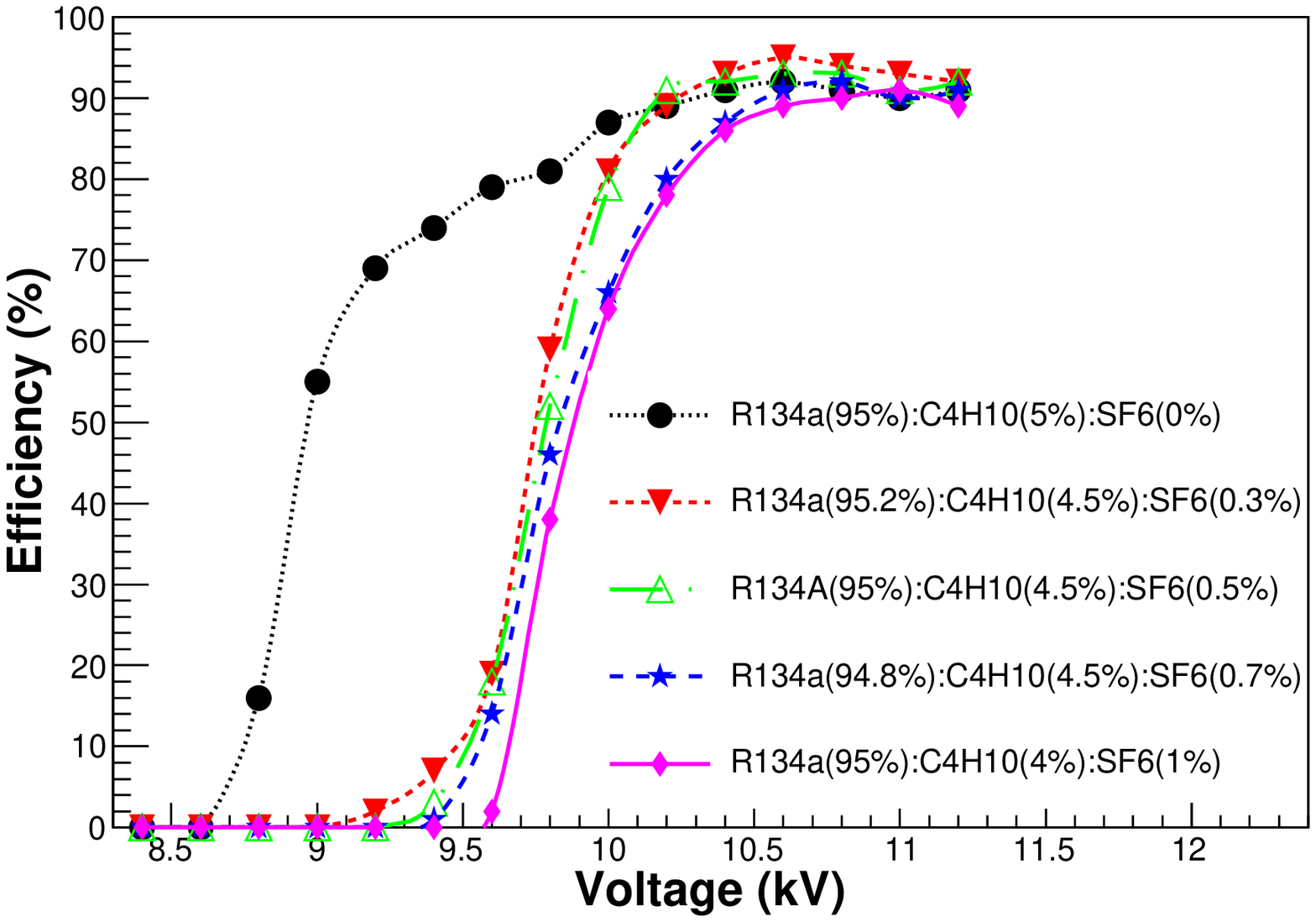}
             \end{minipage}
                    \end{minipage}
\caption{Efficiency of Saint Gobain (left) and Asahi (right) RPC for all the studied five gas mixtures.}
\label{fig:efficiency}
          
 \end{figure}

To determine the optimal gaseous mixture for the ICAL RPC detector, the performance of prototype RPCs was evaluated for a range of gas mixtures.  Single gap RPCs were operated in avalanche mode, using five different gas mixtures, primarily varying the $\rm{SF}_6$ fraction:
{\begin{enumerate}
\item First Mixture : $R134a$ (95.0\%), $C_{4}H_{10}$ (5.0\%), $SF_{6}$ (0.0\%).
\item Second Mixture: $R134a$ (95.0\%), $C_{4}H_{10}$ (4.7\%), $SF_{6}$ (0.3\%).
\item Third Mixture: $R134a$ (95.0\%), $C_{4}H_{10}$ (4.5\%), $SF_{6}$ (0.5\%).
\item Fourth Mixture : $R134a$ (95.0\%), $C_{4}H_{10}$ (4.3\%), $SF_{6}$ (0.7\%).
\item Fifth Mixture : $R134a$ (95.0\%), $C_{4}H_{10}$ (4\%), $SF_{6}$ (1\%).
\end{enumerate}}
The performance of the RPCs with respect to leakage current, count rate, and efficiency was measured, as a function of the applied voltage, for each gas mixture.
Detectors were build from both Asahi and Saint-Gobain glass, and the effect of the type of glass used was also measured.  The results of these performance measurements are shown in Figs.~\ref{fig:crnt}-\ref{fig:efficiency}.  During the course of the experiment the relative humidity varied between 35-40$\%$ and the temperature varied between  $20^{\circ}$C and  $22^{\circ}$C.

Across all five gas mixtures, the leakage current varies between 20 and $80~\rm{nA}$ for RPC built from Saint-Gobain glass. The gas mixture with maximum $SF_6$ concentration (1\%) has the lowest leakage current and the lowest count rate. The error on the current measurement is 2\% of mean value $\pm 9~nA$. The count rate is maximal, approximately $5~Hz/cm^2$, for a gas mixture with no $SF_6$.  The efficiency also decreases with the increase of the $SF_6$ concentration. In case of 0\% $SF_{6}$, the efficiency curve turns on at  $8.8~\rm{kV}$ while at higher concentration it turns on at higher values. However, at the operating region near 10.6 kV, all gas mixtures are more than 90\% efficient. The RPC built with Asahi glass shows similar characteristics with a slight increase in leakage current and count rate at the cost of delayed turn-on for the efficiency.  These results are qualitatively as expected, as the charge suppression from the addition of $\rm{SF}_6$ is expected to reduce leakage current and spurious counts with the slight reduction in efficiency at low voltages. The dataset used for estimating efficiency and count rate is quite large, with statistical uncertainty of the order of a percent, and are within the markers size of the related figures. 

%\begin{figure}[htbp]
%\begin{minipage}{\linewidth}
%      \centering
 %     \begin{minipage}{0.3\linewidth}
  %          \includegraphics[width=5cm,height=6cm]{Fig5a.eps}
     %            \end{minipage}
    %    \begin{minipage}{0.3\linewidth}
       %       \includegraphics[width=5cm,height=6cm]{Fig5b.eps}
       %        \end{minipage}
       %     \begin{minipage}{0.3\linewidth}
      %   \centering
         %     \includegraphics[width=5cm,height=6cm]{Fig5c.eps}
         % \end{minipage}
         %   \end{minipage}
         %   \caption{Leakage current, count rate and efficiency for Saint Gobain RPC for all the five gas compositions studied.}
         %   \label{fig:snt}
  % \end{figure}

 % \begin{figure}[htbp]
 % \begin{minipage}{\linewidth}
    %    \centering
    %    \begin{minipage}{0.3\linewidth}
    %        \includegraphics[width=5cm,height=6cm]{Fig6a.eps}
    %             \end{minipage}
    %    \begin{minipage}{0.3\linewidth}
    %          \includegraphics[width=5cm,height=6cm]{Fig6b.eps}
    %           \end{minipage}
    %        \begin{minipage}{0.3\linewidth}
    %     \centering
    %          \includegraphics[width=5cm,height=6cm]{Fig6c.eps}
    %      \end{minipage}
    %        \end{minipage}
    %        \caption{Leakage current, count rate and efficiency for Asahi RPC for all the five gas compositions studied.}
    %        \label{fig:ash}
  % \end{figure}

\section{Timing Resolution and Charge Distribution}
\label{sec:timing}

In order to differentiate upward from downward-traveling neutrinos, the ICAL detector will precisely measure the arrival time at each RPC in order to distinguish which end was hit first.
When sufficiently well instrumented, the timing resolution of an RPC detector is dominated by fluctuations in the time required for the avalanche to reach the readout electronics.  Generally a wider gaseous region increases the amplitude and results in a higher efficiency, more charge and less time walk.  However, a wider gas gap also increases the time needed by the avalanche to reach the resistive plates, and the associated jitter resulting in a degradation of the timing resolution.  Previous studies have shown that the timing resolution of RPCs decreases as the gas gap increases in both the avalanche and streamer mode operation \cite{cerr}. 

\begin{figure}[htbp]
\begin{minipage}{\linewidth}
%      \centering
      \begin{minipage}{0.46\linewidth}
          \includegraphics[width=8cm,height=6cm]{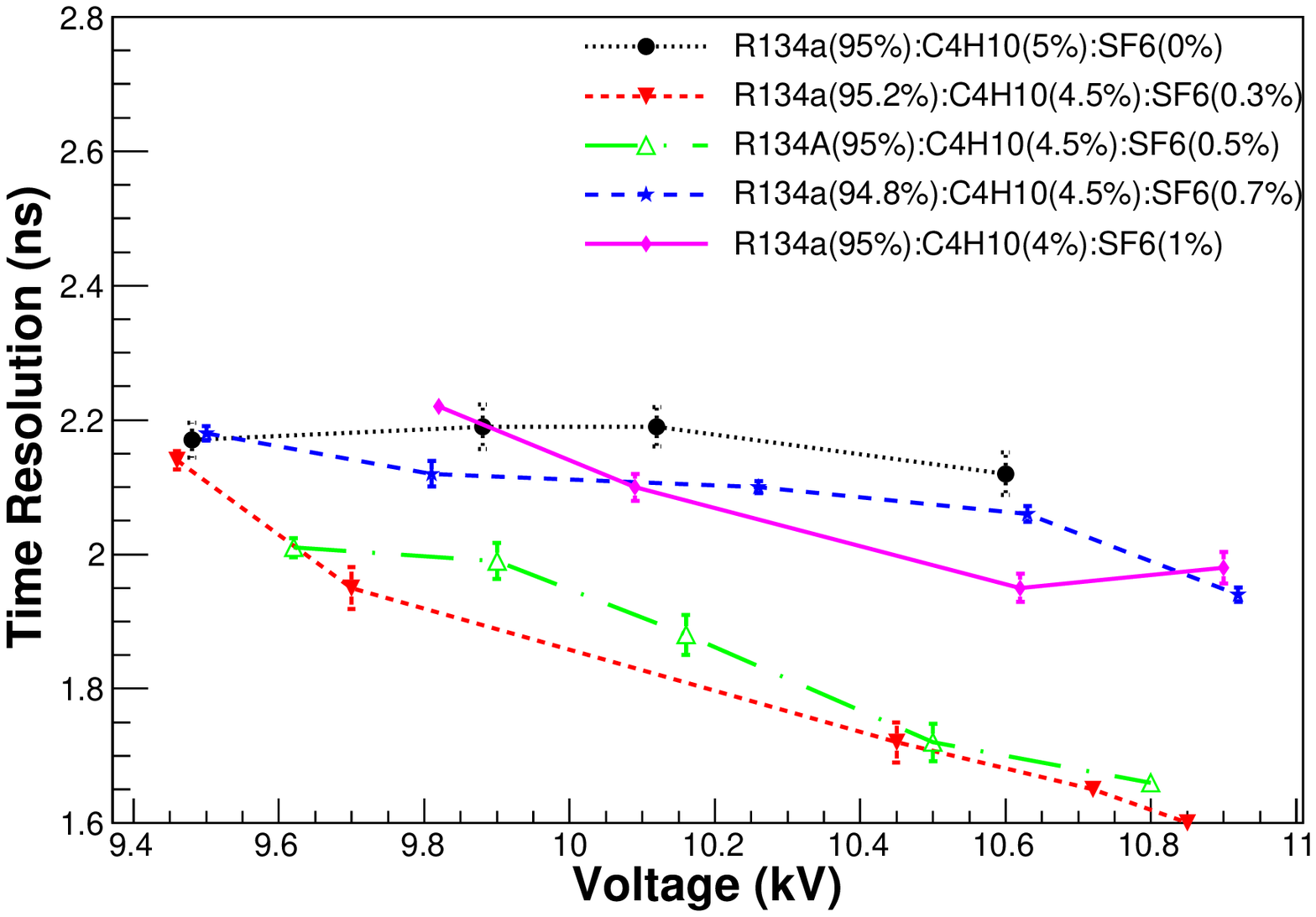}
               \end{minipage}
      \begin{minipage}{0.46\linewidth}
%\centering
            \includegraphics[width=8cm,height=6cm]{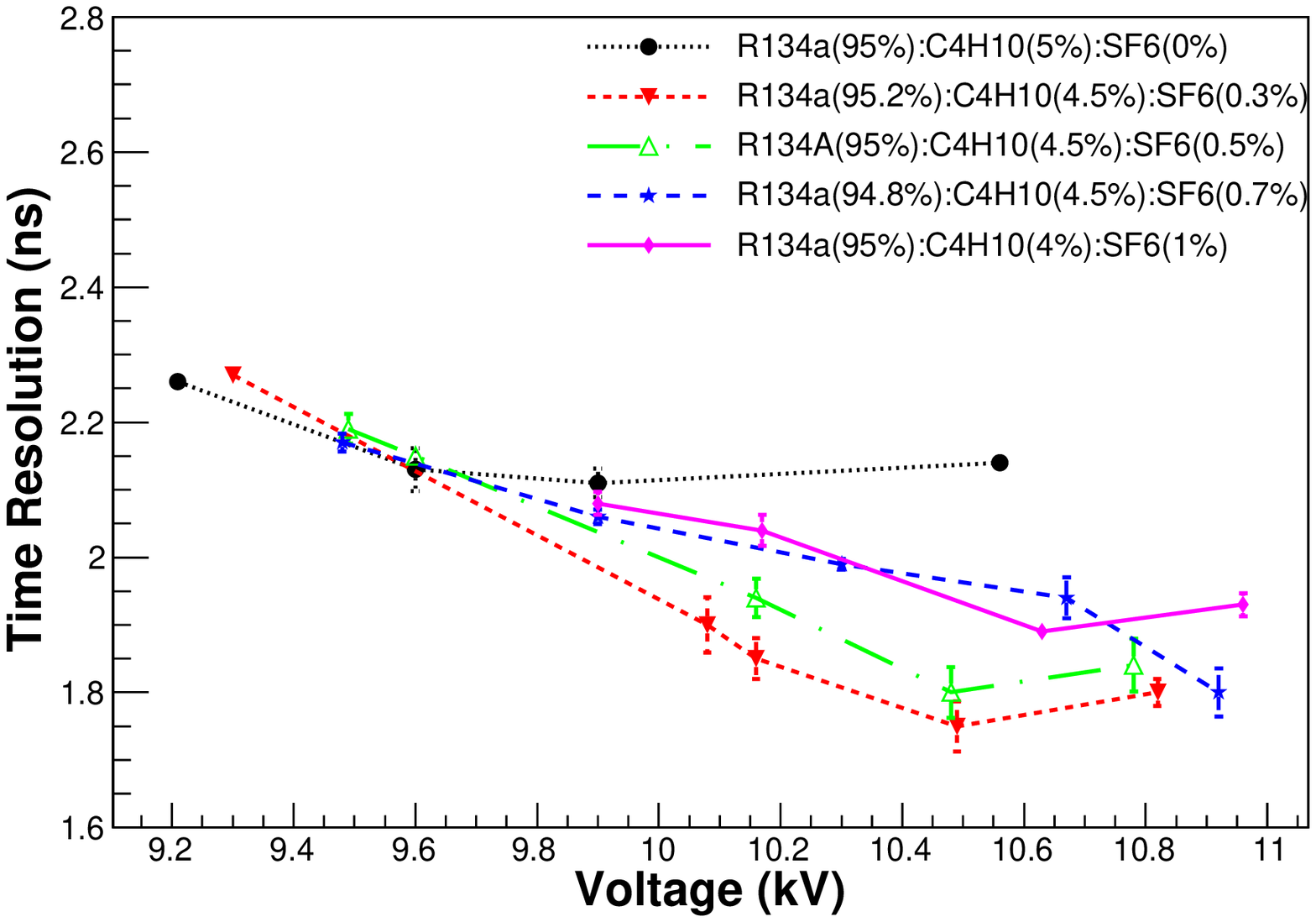}
             \end{minipage}
                    \end{minipage}
\caption{Time resolution of Saint Gobain (left) and Asahi (right) RPC as a function of  effective high voltage.
The time resolution has been estimated from fit to the width of the time distribution and the uncertainty of this is taken as the uncertainty of the  time resolution.}
\label{fig:crct}
          
 \end{figure}

The timing resolution as a function of applied voltage, of prototype RPCs, built with Asahi and Saint-Gobain glass, using a variety of gases, was determined by fitting a Gaussian function to the distribution of STOP-START times, as explained in section 3 and subtracting the estimated contribution from the scintillation counters used in the hodoscope.  The smallest timing resolution is obtained under gas mixture having $\rm{SF}_{6}$ of 0.3\%. For other gas mixtures  there is degradation of time resolution which might be due to the production of less charge along with the deterioration of effective electric field and fluctuations. We measure a timing resolution of $1.6~\rm{ns}$ for Saint-Gobain and $1.7~\rm{ns}$ for Asahi at a $10.6~\rm{kV}$ bias voltage for the gas mixture with $\rm{SF}_6$ fraction of $0.3\%$. Fig.~\ref{fig:crct} shows the corrected timing resolution for both Saint Gobain and Asahi prototypes RPCs. Note, however, that the timing resolution at $\rm{SF}_6$ concentration of 0.5\% is very close to that of 0.3\% $\rm{SF}_6$.

\begin{figure}[htbp]
\begin{minipage}{\linewidth}
      \centering
      \begin{minipage}{0.47\linewidth}
          \includegraphics[width=8cm,height=6cm]{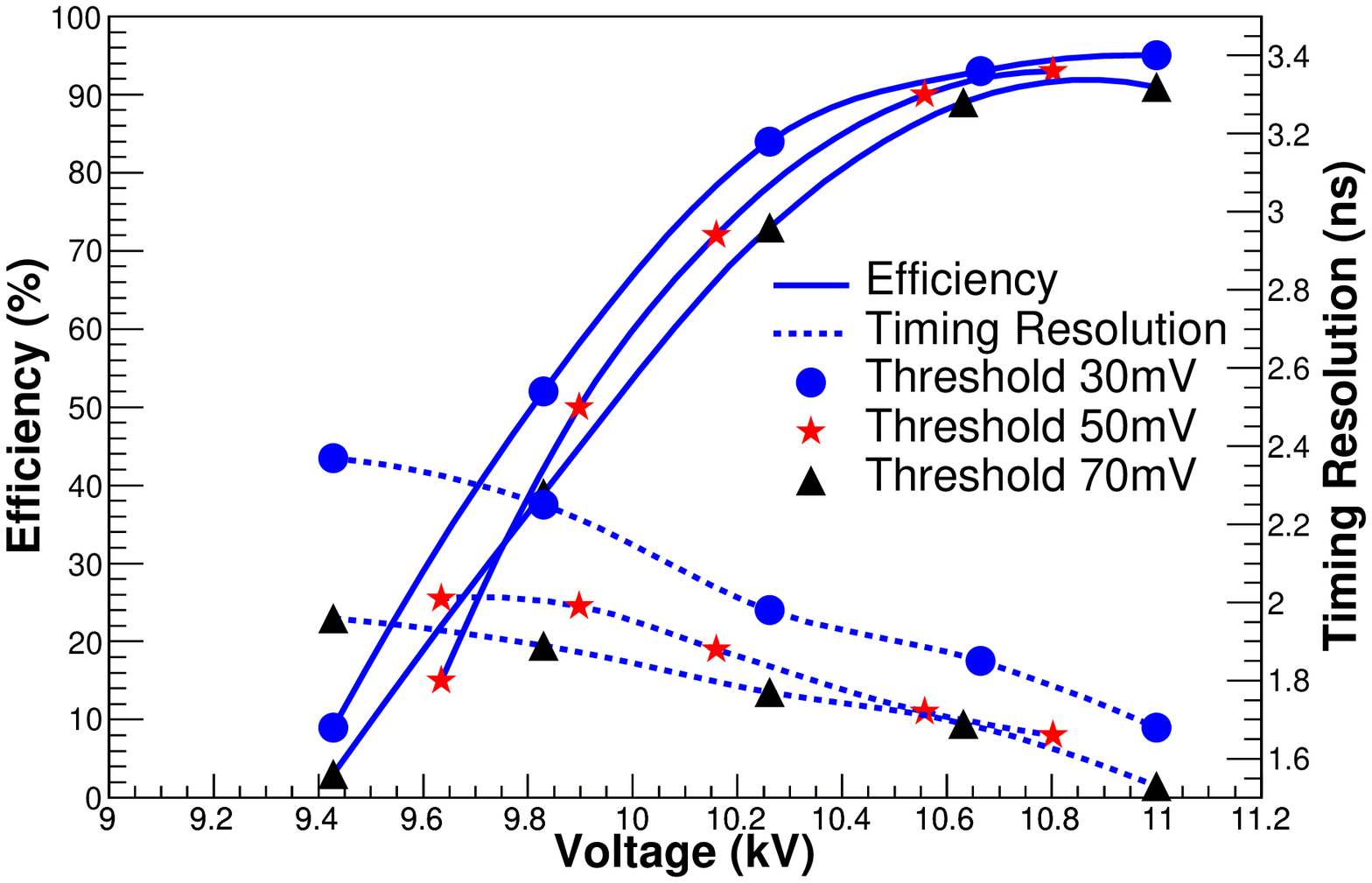}
               \end{minipage}
      \begin{minipage}{0.47\linewidth}
%\centering
            \includegraphics[width=8cm,height=6cm]{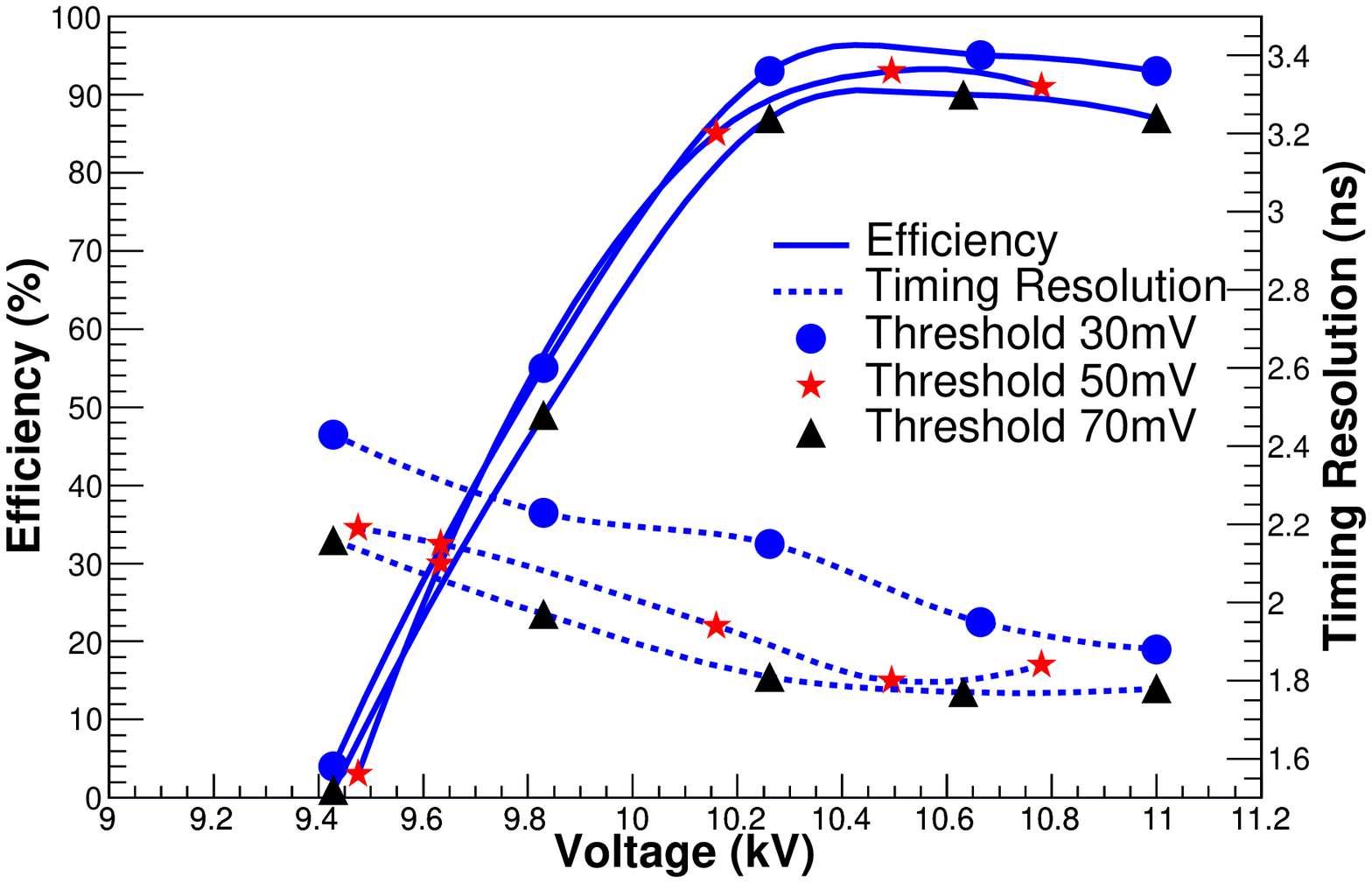}
             \end{minipage}
                    \end{minipage}
\caption{ Timing Resolution and efficiency of Saint Gobain (left) and Asahi (right) as function of effective high voltage  for different discriminator threshold values under the gas mixture $R134a$ (95.0\%), $C_{4}H_{10}$ (4.5\%), $SF_{6}$ (0.5\%).}
          \label{fig:thrs}
          
 \end{figure}

The effect of varying the discriminator threshold value for the output detector pulse on the timing pulse was also determined. The threshold was varied between $30$ and $70~\rm{mV}$ and for third gas mixture i.e., $R134a$ (95.0\%), $C_{4}H_{10}$ (4.5\%), $SF_{6}$ (0.5\%). The timing resolution improves with increase in the threshold value. However, at higher bias voltages, the timing resolution does not vary significantly between the 50 and 70 mV thresholds. Fig.~\ref{fig:thrs} shows the timing resolution of prototype RPCs as a function of applied voltage at various discriminator thresholds under third gas mixture. The timing resolution varies widely at lower bias voltages while at higher voltages the variation is much smaller.  A discriminator threshold of 70 mV gives best timing resolution, but at the cost of a degraded efficiency.
     
\begin{figure}[htbp]
\begin{minipage}{\linewidth}
      \centering
      \begin{minipage}{0.49\linewidth}
          \includegraphics[width=9cm,height=6cm]{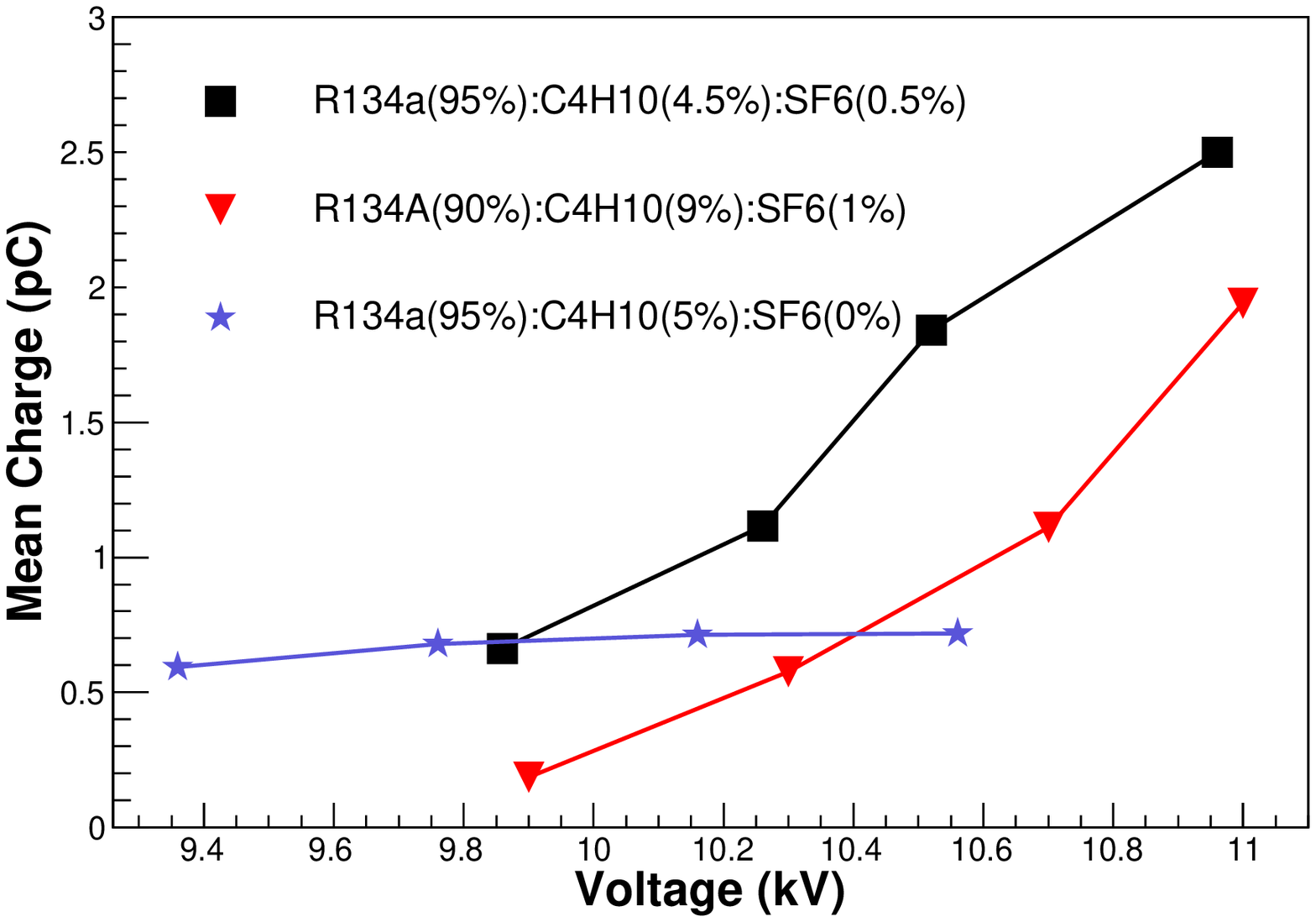}
               \end{minipage}
      \begin{minipage}{0.49\linewidth}
%\centering
            \includegraphics[width=9cm,height=6cm]{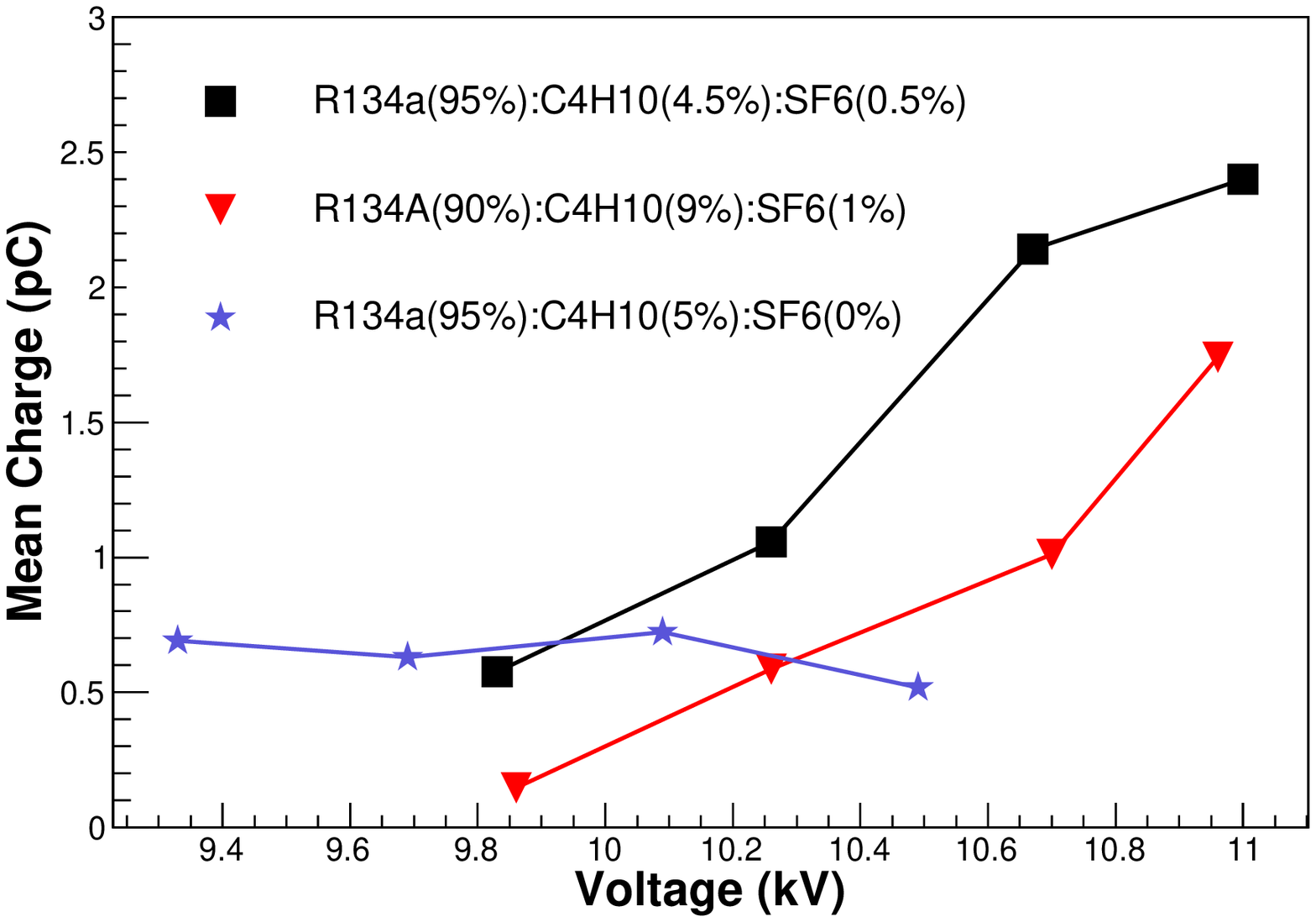}
             \end{minipage}
                    \end{minipage}
\caption{The mean of charge distribution for Saint Gobain (left) and Asahi (right) as a function of  effective high voltage under different gas mixtures.}
          \label{fig:chrg}
          
 \end{figure}

\begin{figure}[htb]
\begin{minipage}{\linewidth}
      \centering
      \begin{minipage}{0.49\linewidth}
          \includegraphics[width=9cm,height=6cm]{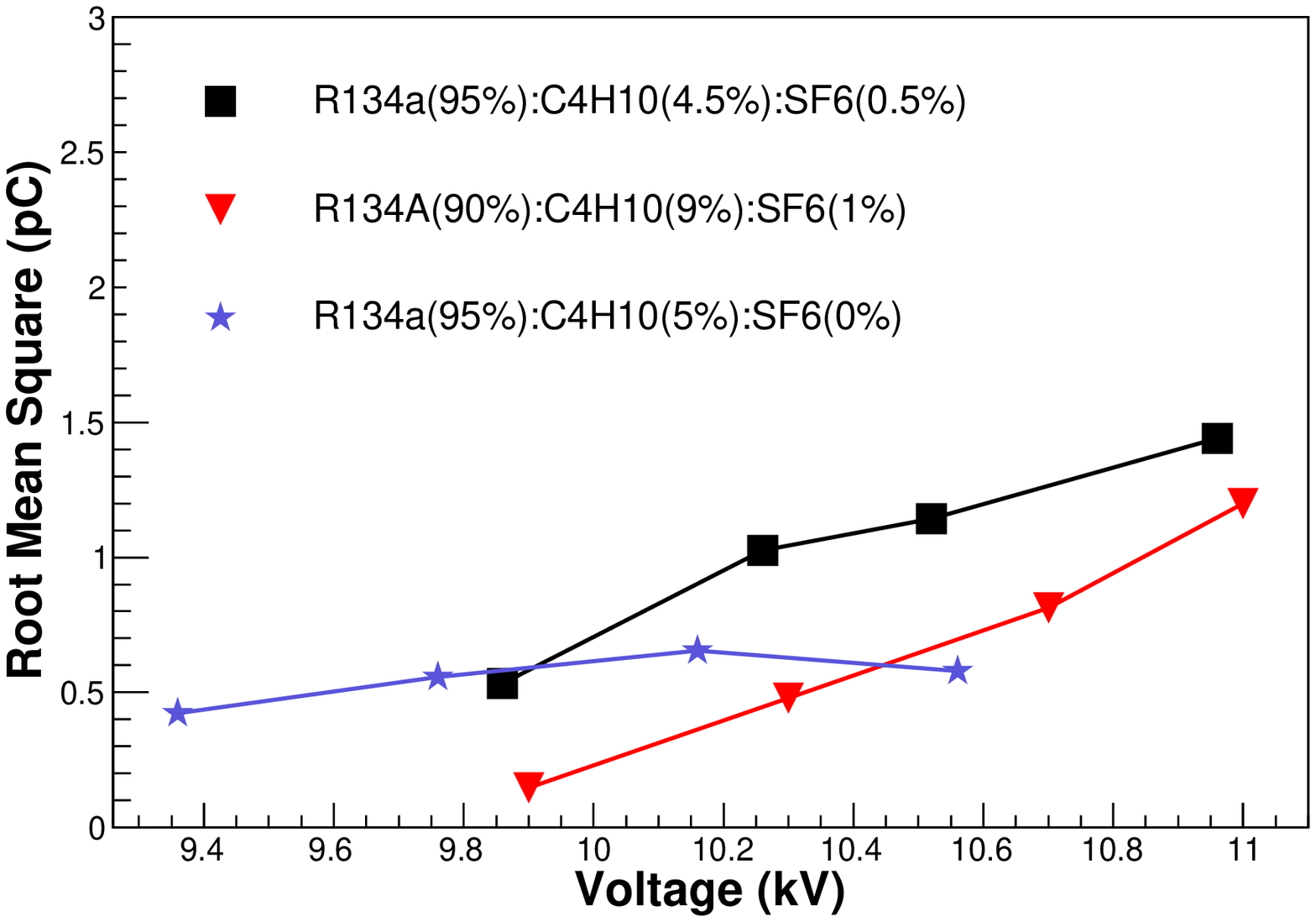}
               \end{minipage}
      \begin{minipage}{0.49\linewidth}
%\centering
            \includegraphics[width=9cm,height=6cm]{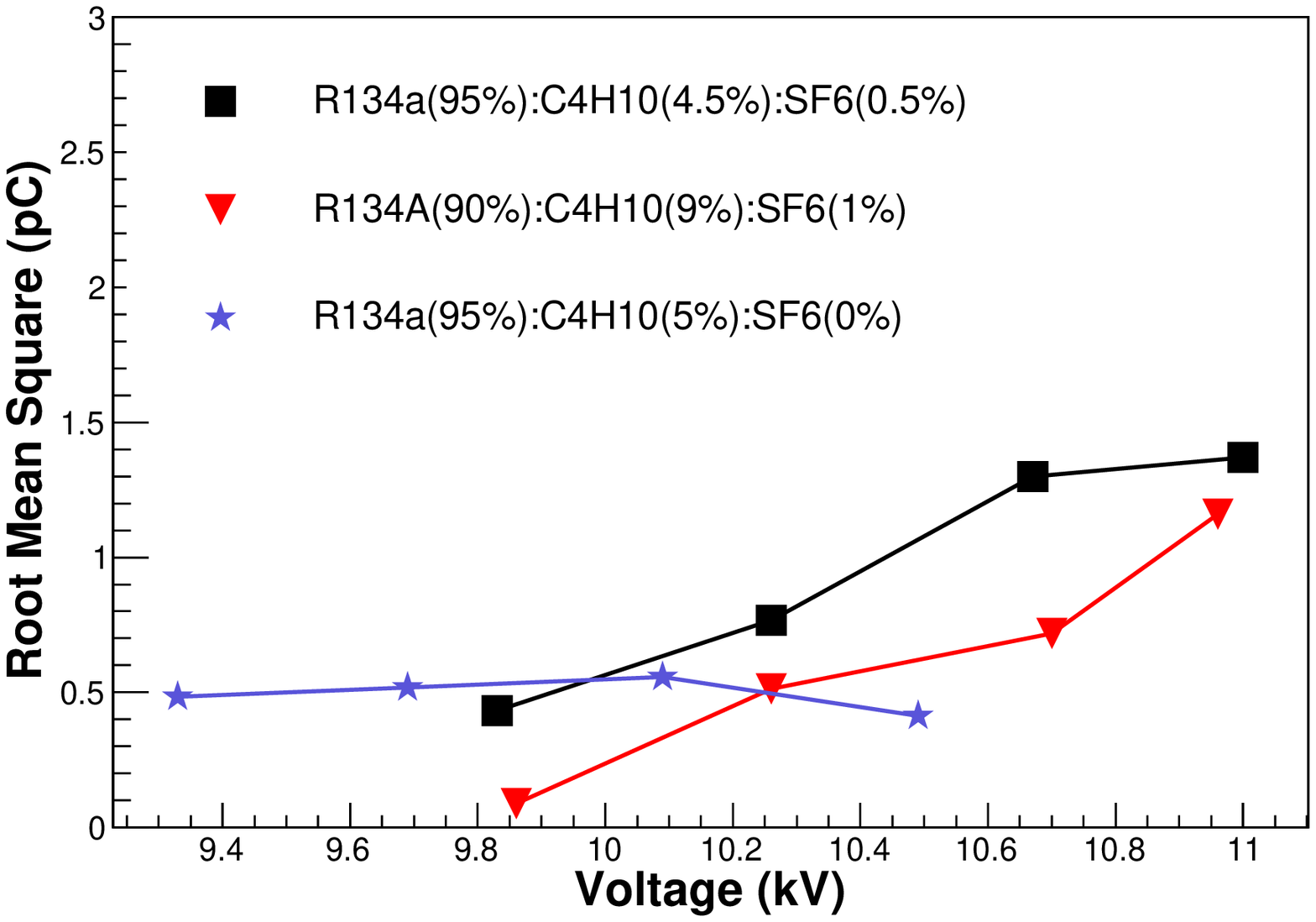}
             \end{minipage}
                    \end{minipage}
\caption{The root mean square distribution of the charge spectra for Saint Gobain (left) and Asahi (right) as a function of  effective high voltage under different gas mixtures.}
          \label{fig:rms}
          
 \end{figure}

The charge content and distribution of output pulses from the RPC were collected using the QDC as described above. The triple coincidence pulse (40 $ns$) of the three scintillators was used as a gate signal for the QDC and the amplified analog pulse from RPC (including delay) was used as an input. The charge output as a function of applied voltage for different prototypes is shown in Fig.~\ref{fig:chrg}. Fig.~\ref{fig:rms} shows the root mean square values of the charge for the Saint Gobain and Asahi RPC as a function of applied voltage.

\section{Conclusions}

The INO facilities ICAL experiment received final approval from the Indian government in early 2015.  The preparations for construction of the tunnel and cavern are currently underway. The collaboration is  set to begin construction of  28,000 RPC detectors of 2 m $\times$ 2 m in size. It is important, at this point, to optimize all the operating parameters of these RPC detectors. The studies reported in this paper are a first attempt to perform a comprehensive characterization of timing and charge measurements for the $3~\rm{mm}$ Asahi and Saint-Gobain glass RPC detectors, which are the final candidates for the first modules of ICAL detector . We found these RPCs to have more than $90\%$ efficiency under all gas mixtures that we studied. The count rate and leakage current are found to be within reasonable limits. The gas mixtures with 0.5\% and 0.3\% $SF_6$ concentration provides comparable time resolutions, which are best amongst all the gas mixtures  studied.  We measured the timing resolution at $1.6~\rm{ns}$ for Saint-Gobain RPC and $1.7~\rm{ns}$ for Asahi RPCs at 0.3\% $SF_6$ concentration and at operating bias voltage of 10.6 kV. The discriminator threshold studies shows that threshold of 30 and 50 mV gives comparable timing resolution and better efficiency compared to a 70 mV threshold at which slight degradation of efficiency is observed . 

The charge spectra shows peak charge collection between 0.5 and 2.5 $pC$ for all the gas mixtures studied. The charge output is nearly constant with bias voltage in the absence of $SF_6$, but it increases with bias voltage with the addition of $SF_6$ concentration in the gas mixtures. The $SF_6$ concentration of 0.5\% results in a mean charge of approximately 0.6 $pC$ and 2.5 $pC$ at 9.8 kV and 11 kV bias voltage respectively. The behavior of charge measurement at 0\% $SF_6$ concentration needs to be further investigated and better understood.

In conclusion, $SF_6$ concentration of 0.3\% provides the best timing resolution while maintaining maximum efficiency, but it suffers from relatively larger count rate and current compared to $SF_6$ concentration of 0.5\%. Moreover, timing resolution and efficiency for $SF_6$ concentration of 0.5\% is comparable to that of 0.3\% $SF_6$ mixture. Since 0.5\% $SF_6$ concentration gives lower count rate, therefore this gas composition is recommended for the INO RPC detectors.  

\section{Acknowledgments}

We would like to thank the Department of Science and Technology (DST), India for generous financial support. We would also like to thank the University of Delhi for providing R\&D grants which have been extremely helpful in the completion of these studies. Our sincere thanks are also due to Prof. Michael Mulhearn of University of California at Davis for proof reading our article.

\end{document}